\documentclass{desyproc}
\begin{document}
\title{Results on Heavy-Flavour Production in pp, p--Pb and Pb--Pb Collisions with ALICE at the LHC}
\author{{\slshape Grazia Luparello$^1$ for the ALICE Collaboration}\\[1ex]
$^1$Universit\`a di Trieste and INFN - Trieste, via A. Valerio 2, Trieste, Italy} 

\contribID{xy}

\confID{8648}  
\desyproc{DESY-PROC-2014-04}
\acronym{PANIC14} 
\doi  

\maketitle
\begin{abstract}
The ALICE Collaboration has measured heavy-flavour production through the reconstruction of hadronic decays of D mesons at mid-rapidity and via semi-electronic (at mid-rapidity) and semi-muonic (at forward rapidity) decays of charm and beauty hadrons in pp, p--Pb and Pb--Pb collisions. 
A summary of the most recent results from p--Pb collisions at $\sqrt{s_{NN}}=5.02$~TeV and Pb--Pb collisions at $\sqrt{s_{NN}} = 2.76$~TeV is presented in this paper.
\end{abstract}
\section{Introduction}
Heavy quarks are effective probes of the Quark Gluon Plasma (QGP) formed in high-energy nucleus-nucleus collisions, since they are produced on a short time scale with respect to that of the QGP. They traverse the strongly interacting medium and lose energy through radiative~\cite{radiativeEloss1} and collisional~\cite{collisionalEloss} processes. 
Theoretical calculations predict a dependence of the energy loss on the colour charge and on the mass of the parton traversing the medium, resulting in a hierarchy in the energy loss with beauty quarks losing less energy than charm quarks, and charm quarks losing less energy than light quarks and gluons~\cite{Dokshitzer,Armesto}.
The energy loss is experimentally investigated via the nuclear modification factor $R_{\mathrm{AA}}$, defined as the ratio of the yield in nucleus-nucleus collisions to that observed in pp collisions scaled by the number of binary nucleon-nucleon collisions.
In the absence of medium effects, $R_{\mathrm{AA}}$ is expected to be unity for heavy flavours, since the production yields are proportional to the number of binary nucleon-nucleon collisions. 
The expected hierarchy in the energy loss described above can be verified comparing the $R_{\mathrm{AA}}$ of different particle species, namely $R_{\mathrm{AA}}(\mathrm B) > R_{\mathrm{AA}}(\mathrm D) > R_{\mathrm{AA}}(light)$. For this comparison it should be considered that the $R_{\mathrm{AA}}$ of the different hadronic species are also affected by the different production kinematics and fragmentation function of gluons, light and heavy quarks.
The $R_{\mathrm{AA}}$ can be modified also due to initial-state effects, since the nuclear environment affects the quark and gluon distributions as described either by calculations based on phenomenological modifications of the Parton Distribution Functions (PDF)~\cite{Eskola} or by the Colour Glass Condensate (CGC) effective theory~\cite{Fujii}. Partons can also lose energy in the initial stages of the collision via initial-state radiation~\cite{Vitev}, or they can experience transverse momentum broadening due to multiple soft collisions prior to the hard scattering~\cite{Lev}. Initial-state effects are addressed by studying p--Pb collisions.
Finally, in nucleus-nucleus collisions the charmed hadron azimuthal anisotropy, quantified via the second order coefficient of the Fourier decomposition of the particle momentum azimuthal distribution ($v_2$), tests whether also charm quarks participate in the collective expansion dynamics and possibly thermalize in the QGP. 
\section{Open heavy-flavour measurements in ALICE}
The excellent performance of the ALICE detector~\cite{Abelev:2014ffa} allows open heavy-flavour measurements in several decay channels and in a wide rapidity range. 
At mid-rapidity ($|y|<0.5$) D mesons are reconstructed via their hadronic decay channels: $\mathrm {D^0} \rightarrow \mathrm {K^-} \pi^+$, $\mathrm{D^+} \rightarrow \mathrm{K^-} \pi^+ \pi^+$, $\mathrm{D^{*+}} \rightarrow \mathrm{D^0} \pi^+ \rightarrow \mathrm{K^-} \pi^+ \pi^-$, $\mathrm{D_{\mathrm s}^+} \rightarrow \phi  \pi^+ \rightarrow  \mathrm{K^-} \mathrm{K^+} \pi^+$ and their charge conjugates. D-meson selection is based on the reconstruction of decay vertices displaced by a few hundred $\mu$m from the interaction vertex, exploiting the high track-position resolution close to the interaction vertex provided by the Inner Tracking System (ITS). The large combinatorial background is reduced by selections applied on the decay topology and by the identification of charged kaons and pions in the Time Projection Chamber (TPC) and Time-Of-Flight (TOF) detector. 
Electrons are identified at mid-rapidity through their specific energy loss in the TPC gas combined with the information from the TOF and from the electromagnetic calorimeter (EMCal). 
At forward rapidity, open heavy-flavour production is studied in the semi-muonic decay channel. Muons are reconstructed in the five tracking stations of the Muon Spectrometer ($-4 < \eta < -2.5$). The reconstructed tracks are matched with tracklets measured in the trigger stations to reject punch-through hadrons. 
\section{Results} 
\begin{figure}[htb]
\centerline{\includegraphics[width=0.45\textwidth]{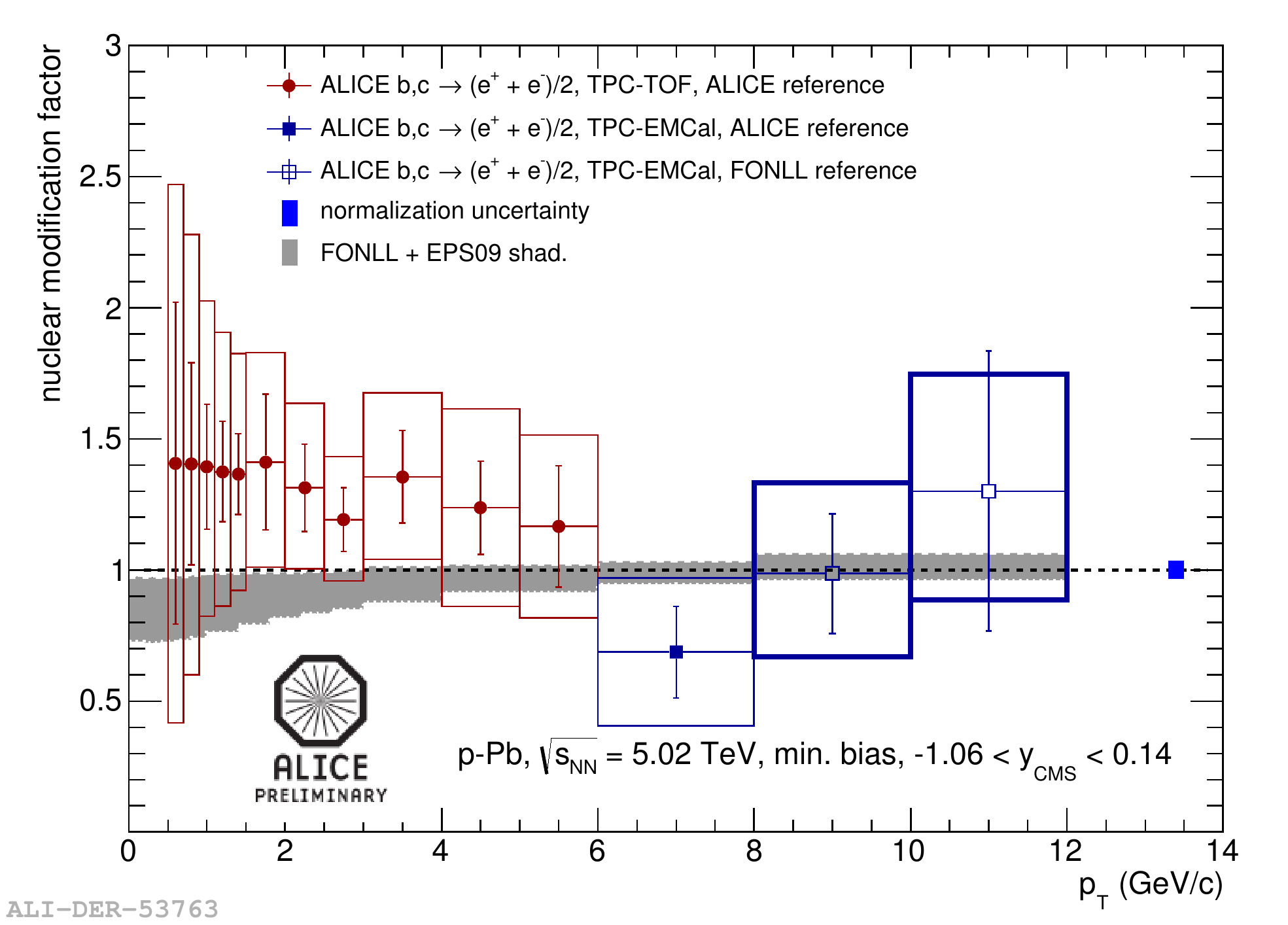}
\includegraphics[width=0.48\textwidth]{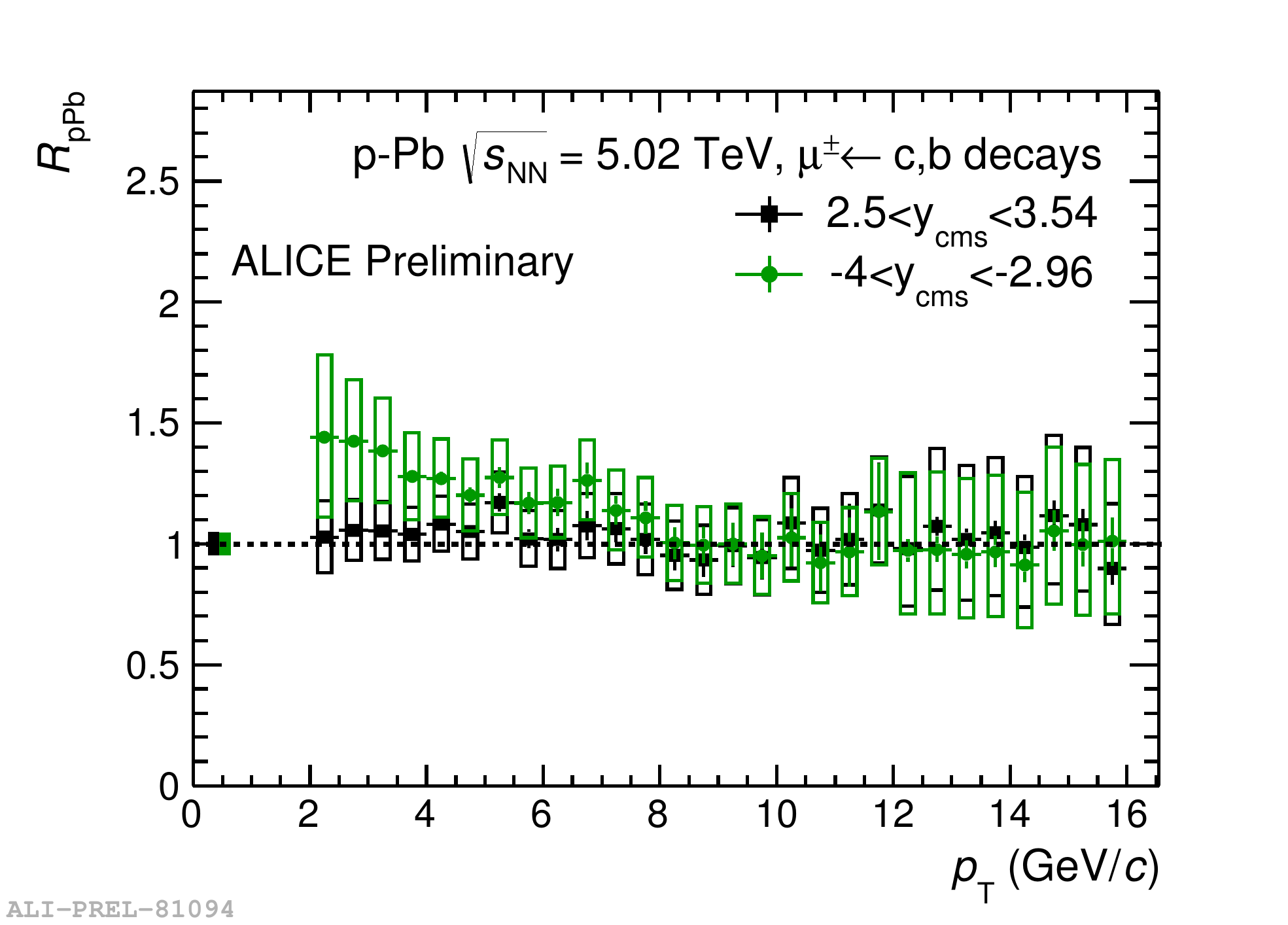}}
\vspace{-10pt}
\caption{$R_{\mathrm{pPb}}$ of heavy-flavour decay electrons at mid-rapidity and heavy-flavour decay muons at forward (p-going direction) and backward (Pb-going direction) rapidities in minimum-bias p--Pb collisions, as a function of $p_{\mathrm T}$.}\label{Fig:RpA}
\end{figure}

Figure~\ref{Fig:RpA} shows the nuclear modification factor ($R_{\mathrm{pPb}}$) measured in p--Pb collisions at $\sqrt{s_{\mathrm{NN}}}$=5.02 TeV as a function of $p_{\mathrm T}$ for heavy-flavour decay electrons (left) and muons (right). 
The measurement of D-meson $R_{\mathrm{pPb}}$ is reported in~\cite{DmesonRpA}.
The results are compatible with unity within uncertainties without any significant dependence on the rapidity interval investigated. The measurements confirm that initial-state effects due to the presence of cold-nuclear matter are small in the measured $p_{\mathrm T}$ range. Theoretical predictions based on pQCD calculations including the EPS09~\cite{Eskola} nuclear modification of the PDF can describe the measurements. The D-meson $R_{\mathrm{pPb}}$ is also compatible with calculations based on the CGC~\cite{Fujii} and with a model including cold-nuclear-matter energy loss, nuclear shadowing and $k_{\mathrm T}$-broadening~\cite{Sharma}. 

\begin{figure}[htb]
\centerline{
\includegraphics[width=0.38\textwidth]{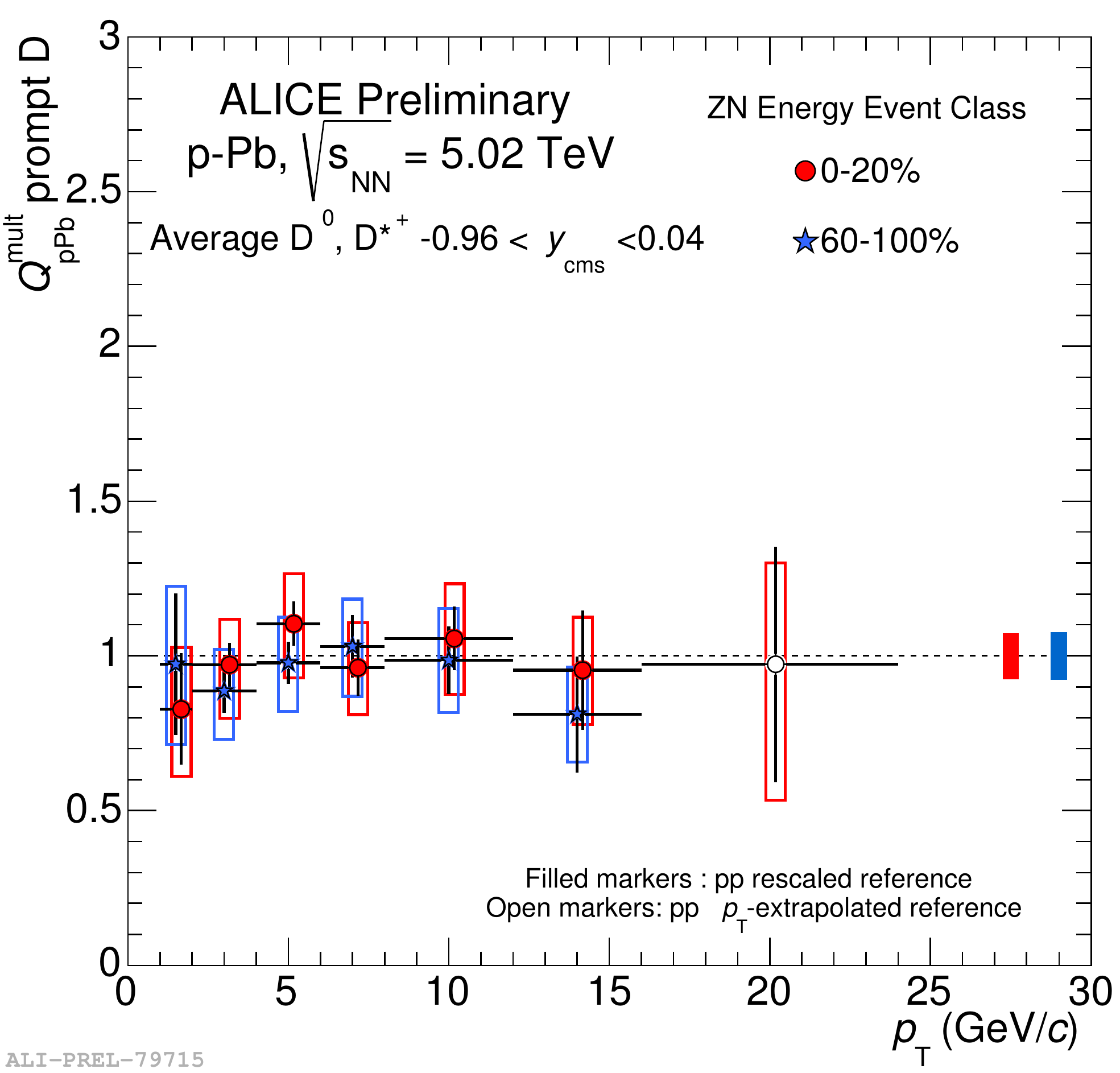}}
\vspace{-10pt}
\caption{Prompt D-meson $Q_{\mathrm{pPb}}^{\mathrm{mult}}$ in the 0-20\% and 60-100\% ZNA energy classes.}\label{Fig:DVsMult}
\end{figure}

Open heavy-flavour production is also studied in p--Pb collisions as a function of the event activity.
The ratio $Q_{\mathrm{pPb}}^{\mathrm{mult}}(p_{\mathrm T}) = \frac{{\mathrm d} N_{\mathrm{pPb}}^{\mathrm{mult}}/{\mathrm d} p_{\mathrm T}}{<N_{\mathrm{coll}}^{\mathrm{mult}}> {\mathrm d} N_{\mathrm{pp}}/{\mathrm d} p_{\mathrm T}}$
is used to study the possible multiplicity-dependent modification of the $p_{T}$-differential yields in p--Pb collisions with respect to the binary-scaled yields measured in pp collisions. 
Events are divided in classes based on the energy measured in the Zero Degree Calorimeters located in the Pb-going direction (ZNA). The average number of nucleon-nucleon collisions for the considered ZNA energy event class, $<N_{\mathrm{coll}}^{\mathrm{mult}}>$, is calculated with the hybrid approach described in~\cite{Toia}. Figure~\ref{Fig:DVsMult} demonstrates that the $Q_{\mathrm{pPb}}^{\mathrm{mult}}$ of prompt D mesons for events with high and low multiplicities is compatible with unity within uncertainties, thus no multiplicity-dependent modification of D-meson production in p--Pb collisions relative to the binary-scaled pp production is observed. 

\begin{figure}[htb]
\centerline{\includegraphics[width=0.38\textwidth]{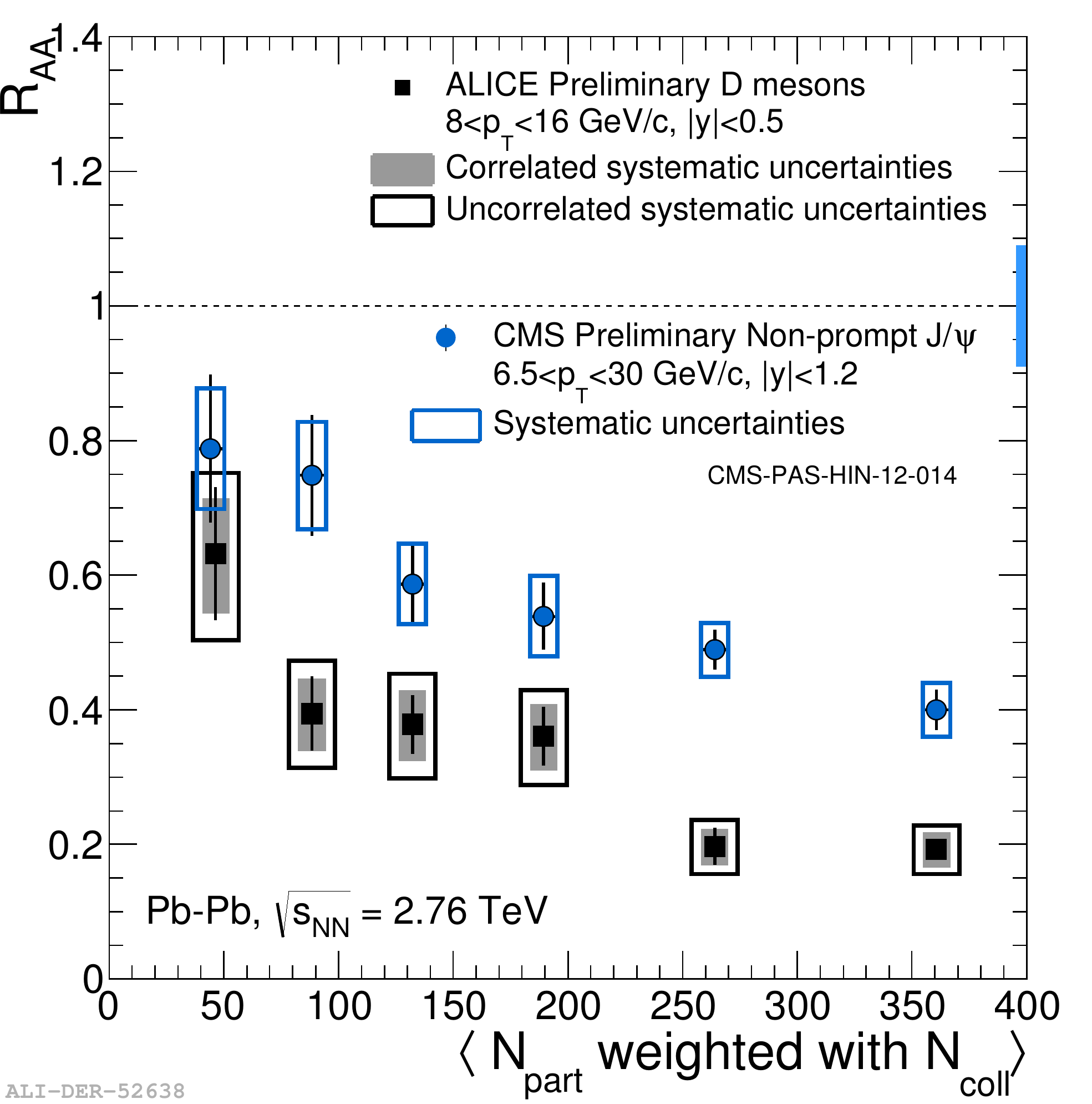} 
\includegraphics[width=0.42\textwidth]{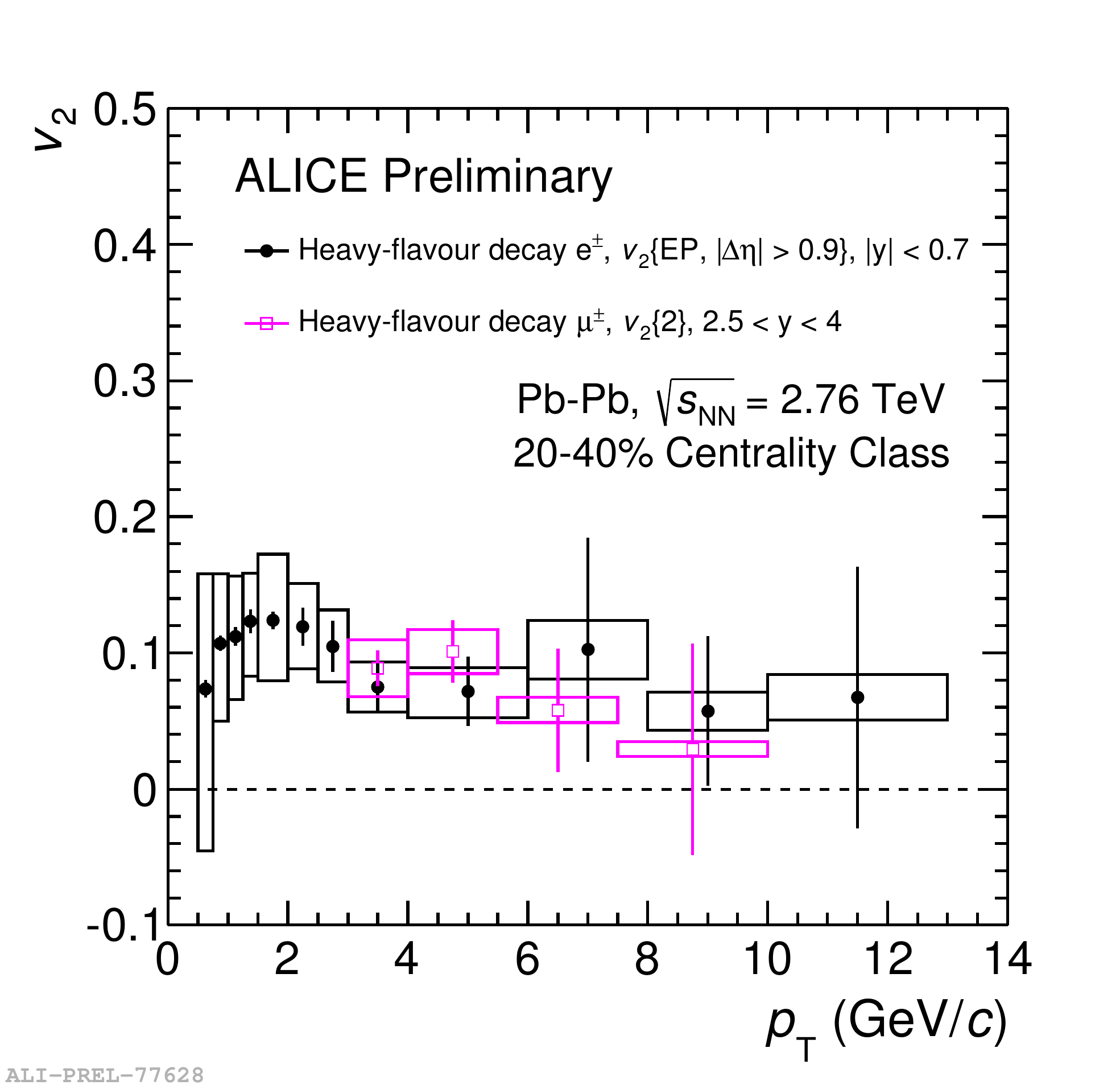}}
\vspace{-10pt}
\caption{Left: $R_{\mathrm{AA}}$ of prompt D mesons and of non-prompt J/$\psi$ measured by CMS~\cite{CMS} as a function of centrality, expressed in terms of the number of nucleons participating in the interaction. Right: Heavy-flavour decay electron and muon $v_2$.}\label{Fig:PbPb}
\end{figure}

In Pb--Pb collisions, the open heavy-flavour $R_{\mathrm{AA}}$ measured with ALICE in the different channels~\cite{RAA} shows a strong reduction of the yields at large trasverse momenta ($p_{\mathrm T} >5$~GeV/$c$) in the most central collisions relative to a binary-scaled pp reference. This suppression is interpreted as due to charm quark in-medium energy loss. The expected mass ordering of the energy loss is also investigated: Fig.~\ref{Fig:PbPb} (left) shows the D-meson $R_{\mathrm{AA}}$ as a function of centrality, represented as the average number of nucleons participating in the interaction, compared to the one of J/$\psi$ from beauty-hadron decays measured by CMS~\cite{CMS}. The D-meson $p_{\mathrm T}$ range was chosen in order to obtain a significant overlap with the $p_{\mathrm T}$ distribution of B mesons decaying to J/$\psi$ with $6.5 < p_{\mathrm T} < 30$~GeV/$c$, thus allowing a consistent comparison. 
A similar trend as a function of centrality is observed, but the D-meson $R_{\mathrm{AA}}$ is systematically lower than the one of J/$\psi$ from B decays. This is consistent with the expectation of a smaller in-medium energy loss for beauty than for charm quarks. 
A comparison with the $R_{\mathrm{AA}}$ of charged hadrons and pions is also done (not shown): a similar suppression is observed, although the uncertainties do not allow yet to draw a conclusion on the colour-charge dependence of the in-medium energy loss.
Figure~\ref{Fig:PbPb} (right) shows the $v_2$ of heavy-flavour decay electrons and muons in the centrality interval 20-40\%. The two results are compatible within uncertainties. For $2<p_{\mathrm T}<3$~GeV/$c$ a positive $v_2$ is observed ($>3\sigma$ effect).
The D-meson $v_2$ measured in the 30-50\% centrality class is larger than zero with a 5.7$\sigma$ significance in the interval $2<p_{\mathrm T}<6$~GeV/$c$ and comparable in magnitude to the one of charged hadrons~\cite{v2PRL} (not shown). These results indicate that heavy quarks participate in the collective motion of the system. 
At high $p_{\mathrm T}$, $v_2$ results could give insight into the path-length dependence of the in-medium energy loss, but the present statistics does not allow to conclude on this. 

In summary, in p--Pb collisions the open heavy-flavour $R_{\mathrm{pPb}}$ is consistent with unity indicating that initial-state effects are small. In Pb--Pb collisions a large suppression of open heavy-flavour yields is observed at intermediate and high $p_{\mathrm T}$. Since initial-state effects are small, these results can be interpreted as a final-state effect due to the interaction of the charm quarks with the hot and dense medium. The $v_2$ measured in Pb--Pb semi-central collisions is larger than zero at low $p_{\mathrm T}$, suggesting that heavy quarks participate in the collective motion of the system.
 
\begin{footnotesize}

\end{footnotesize}
\end{document}